\documentclass[11pt]{article}
\usepackage{epsf}
\usepackage{pazh}
\usepackage{flushrt}
\usepackage{pstricks}

\def\deg{\hbox{$^\circ$}}

\begin{document}

{\tiny To be published in Astronomy Letters, Vol. 28, No. 12, 2002, pp.
811-820\\ Translated from Pis'ma v Astronomicheskij Zhurnal, Vol. 28, No.
12, 2002, pp. 902-912}

\title{Quasi-Periodic X-ray Oscillations in the Source Sco~X-1}{QPOs in
Sco~X-1}

\author{Sergey~Kuznetsov\footnote{\tt{Email: sik@hea.iki.rssi.ru}}}
\affil{\it \centerline{Space Research Institute, Profsoyuznaya 84/32, Moscow,
117997 Russia}}

\begin{center}
{\small Received: July 5, 2002\\}
\end{center}

\section*{}
{\small The RXTE observations of Scorpius~X-1 in 1996--1999 are
presented. The properties of its quasi-periodic X-ray oscillations are
studied in detail. The results obtained are used for analysis in terms
of the transition-layer model (TLM) and the relativistic-precession
model (RPM) for a slowly rotating neutron star. Theoretical
predictions of the two models are compared and their self-consistency
is verified. The tilt of the magnetosphere to the accretion-disk
plane, the neutron-star mass, and its angular momentum are
determined in the model frameworks.

\vspace*{10pt} \emph{Key words:} accretion, neutron
stars, Scorpius~X-1, X-ray sources, low-mass binary
X-ray sources, quasi-periodic oscillations.}


\section{Introduction}
Scorpius~X-1 is a low-mass X-ray binary with an accreting neutron star
and one of the brightest X-ray sources. Its spectral properties
suggest that Scorpius~X-1 belongs to the class of Z-type sources
(Hasinger and van der Klis~1989), whose characteristic feature is a
Z-shaped track in the color--color diagram. In this interpretation,
the spectral properties are presented in the hard--soft color indices,
each of which is defined as the harder-to-softer flux ratio in the
corresponding energy band. The Z-shaped track is commonly divided into
three parts called branches: the horizontal (HB, the upper part of the
diagram), normal (NB, the intermediate part), and flaring (FB, the
lower part) branches. The position on the Z~track is generally
believed to be associated with the rate of accretion in the direction
from HB to FB. Six sources are currently known to exhibit Z~tracks in
the color--color diagram: Scorpius~X-1, Cygnus~X-2, GX~17+2, GX~5-1,
GX~340+0, and GX~349+2.

The power-density spectra (Fourier transforms of the flux) of Z-type
sources exhibit low-frequency (5--100~Hz) X-ray quasi-periodic
oscillation (QPO) peaks. The names of the QPOs correspond to the
branch with which their origin is identified: horizontal- (HBO),
normal- (NBO), and flaring-branch (FBO) oscillations. HBOs
(15--100~Hz) can also be detected in the NB spectral state. However,
as one recedes from HB, the statistical significance of the QPO peaks
decreases and they become undetectable. When moving along the Z~track
(from HB to FB) in its NB--FB segment, a QPO peak detectable in the
range 5--20~HZ (NBO/FBO) emerges in the power-density spectra. For all
the currently known Z-type sources, QPOs were also found in the range
200--1100~kHz (van der Klis~2000). Two kHz QPO peaks ($\nu_{1}$\ and
$\nu_{2}$ are the lower and upper peaks, respectively) can be
simultaneously observed with a frequency difference of $\sim
200$--400~Hz. For a series of successive observations (i.e., on short
time scales from several hours to several days), an increase in the
flux is accompanied by an increase in the frequencies of the two
peaks, but their difference ($\Delta\nu=\nu_{2}-\nu_{1}$) is not
conserved (van der Klis et al.~1997).

The observations of Scorpius~X-1 revealed all the QPO types
characteristic of the low-frequency ($<100$~Hz) range in the
power-density spectrum. At high frequencies, two kHz QPO peaks
($\nu_{1} and \nu_{2}$) were found in the RXTE/PCA data [first
detected by van der Klis et al.~(1996)]. Here, we study the QPO
properties in detail. Our results are used for analysis in terms of
the transition-layer model (TLM) and the relativistic-precession model
(RPM). We compare theoretical predictions of the two models and verify
their self-consistency. The X-ray flux variability of the source at
low frequencies (0.1--128~Hz) is investigated by taking into account
the power-law behavior of the power-density spectrum at frequencies
above and below the break frequency.

\section{DATA AND OBSERVATIONS}
For our time analysis, we used data from the PCA (Proportional Counter
Array) instrument (Jahoda et al.~1996) onboard the RXTE observatory
(Bradt et al.~1993) retrieved from the Goddard Space Flight Center
Electronic Archive.

The source Scorpius~X-1 was observed by the RXTE observatory during
eleven series of pointing observations (10056, 10057, 10059, 10061,
20053, 20426, 30035, 30036, 30406, 40020, 40706): in February and
May~1996, in March, April, and August~1997, in January, February, and
from May until July~1998, and in January and July~1999. The
observations of Scorpius~X-1 over this period correspond to three
different observational epochs of RXTE/PCA (1, 3, and 4 in the adopted
classification), for which the boundaries of the PCA energy channels
were changed.

To construct the power-density spectra, we used observational data
with a resolution of $\sim$122 or $\sim244$~$\mu$m ($2^{-13}$,
$2^{-12}$~s) from the zeroth to 87th or 249th PCA energy
channels. These ranges (0--87 and 0--249) correspond to a detectable
photon flux $\sim1.5-23.5$, $\sim1.9-32.3$, and $\sim2.0-38$~keV for
epochs 1, 3, and 4, respectively, or extend to $\sim60$~keV if the
entire accessible PCA energy range (channels 0--249) was used.

Of all the observations, we used only those during which the angle
between the source direction and the Earth's horizon was more than
10$\deg$. During the observations of Scorpius~X-1, all five
proportional counters were not always switched on to record events. If
the operating condition of one of the counters changed during a
continuous observation (whose duration did not exceed the duration of
one orbit and was, on average, $3-3.5\times10^3$~s), then the time
interval during which the total count rate changed abruptly was
excluded from our analysis.

To analyze the variability of Scorpius~X-1, we constructed its
power-density spectra (van der Klis~1989) in the frequency range
0.03125--2048~Hz (see Fig.~1). The properties of the low- and
high-frequency flux variability were investigated in the bands
0.1--256 and 256--2048~Hz, respectively. No correction was made for
the background radiation and dead time.

\begin{figure}[htb]
\epsfxsize=9.5cm
\epsffile{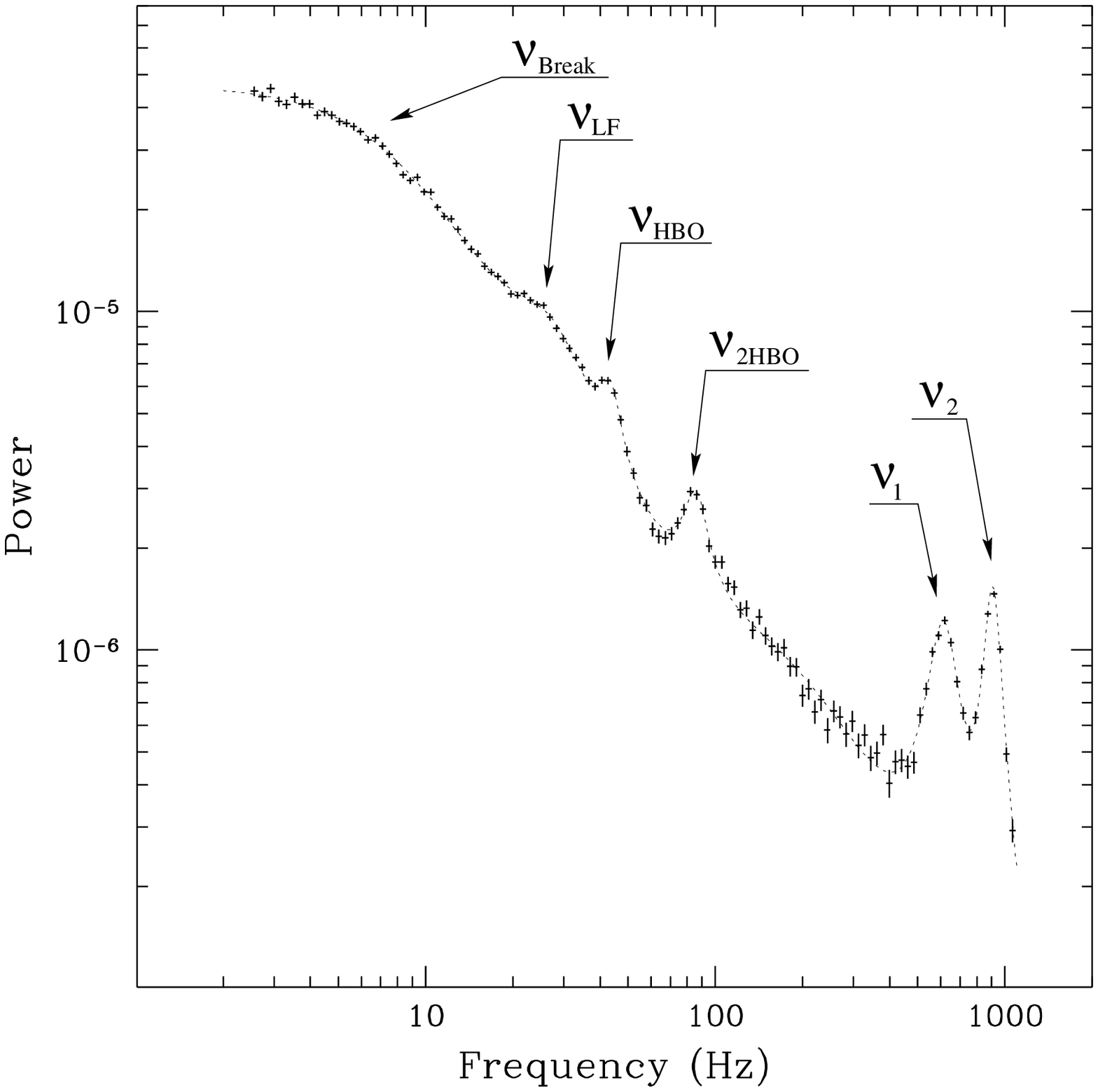}
\rput[tl]{0}(9.5,9.6){
\begin{minipage}{7.2cm}
\normalsize\parindent=0.0mm
{\small {\bf Fig.~1}. The power-density spectrum for the Z-type
source Scorpius~X-1. The data were averaged
over four consecutive observing intervals from
$05:04:13$~UTC on May~25, 1996. The total observing
time is ${\sim 10^{4}}$~s [see a similar
power-density spectrum in Titarchuk et
al.~(1999)]. The X-ray QPO peaks are shown: HBOs
($\nu_{{HBO}}$) and their second harmonic
($\nu_{2{HBO}}$) and the kHz QPO peaks
($\nu_{1}$ and $\nu_{2}$). The break frequency is
denoted by $\nu_{{Break}}$. A model fit to the
data (for more details, see the text) is indicated by
the dotted line. Broadband noise was detected at a
statistically significant level in the average
power-density spectrum at low frequencies
($\nu_{{LF}}$).}
\end{minipage}}
\end{figure}

We analyzed individual observations with a duration up to $\sim
3.5\times 10^{3}$~s. Fitting the power-density spectra by a constant
and by a power law at frequencies below and above the break frequency
did not yield acceptable results (according to the $\chi^{2}$
test). The main reason was the absence of a sharp break and the
resulting uncertainty in its measurement. The model in which at
frequencies much higher ($\nu/\nu_{{break}}
\gg 1$) and much lower ($\nu/\nu_{{break}} \ll
1$) than the break, each part of the spectrum could be fitted by its
own power law and the transition between them was not jumplike
(previously, this model was successfully used for a time analysis of
the Z-type source Cygnus~X-2; see, e.g., Kuznetsov~2001), proved to be
more suitable:
\begin{equation}\label{eq:break:Kuznetsov_n}
P(\nu)=A\nu^{-\alpha}[1+(\nu/\nu_{\textrm{break}})^{\beta}]^{-1} .
\end{equation}
The power-density spectra (see Fig.~1) were fitted by this model in
the $0.1$--$128$-Hz band with the additional introduction of one or
two Lorenz lines to allow for the QPO peaks and their harmonics. To
take into account the PCA dead-time effect, which causes the overall
level to be shifted to the negative region [because of this effect,
the Poissonian noise level subtracted from all spectra differed from
$2.0$, in units of the Leahy normalization; see van der Klis~(1989) and
Vikhlinin et al.~(1994) for more details], we added a constant to the
general model.

In searching for the kHz (in the range $\sim500-1200$~Hz) QPO peaks
$\nu_{1}$ and $\nu_{2}$, we analyzed the power-density spectra at high
frequencies. A constant with the addition of one or two Lorenz lines
was used as the model. Individual observations in which the detection
confidence level of kHz QPO peaks exceeded $3\sigma$ were used for the
subsequent analysis.

Our analysis of the low-frequency flux variability in Scorpius~X-1
revealed QPOs at a high confidence level in all the individual
observations that corresponded to the HB or NB spectral states.  Note
that the HBO frequency is one of the main frequencies in most models,
including the TLM and RPM.  The two models (see below) unequivocally
establish the correspondence between the three QPO peaks: the HBOs,
lower, and upper kHz QPOs.

To obtain the most accurate model parameters (the angle $\delta$ for
the TLM and the neutron-star mass $M_{{NS}}$ and angular momentum $a$
for the RPM), which are invariants in each of the theories under
consideration, it was necessary to calculate the three simultaneously
observed main QPO frequencies (i.e., $\nu_{{HBO}}, \nu_{1}, \nu_{2}$;
see Fig.~1) as accurately as possible. Therefore, of all the
preselected data in which both kHz QPO peaks ($\nu_{1}$ and $\nu_{2}$)
were found, we used only those in which the main HBO peak
($\nu_{{HBO}}$) and its second harmonic ($\nu_{2{HBO}}$) were detected
at a confidence level above $4\sigma$. In such observations, the two
kHz QPO peaks were also reliably detected.

\section{LOW-FREQUENCY X-RAY FLUX VARIABILITY}

\subsection{The Power of the HBO peaks}

In Fig.~2, rms is plotted against HBO frequency. The flux variability
(i.e., $rms^{2}$) at the peak corresponds to the integral of the
Lorenz line that fits best the QPO peak in the power-density spectrum.
For the Z-type sources, to which Scorpius~X-1 belongs, an increase in
HBO frequency is accompanied by a decrease in rms (see, e.g., van der
Klis~2000). As a result, the detection confidence level of HBO peaks
is lower at high frequencies (${\sim50-60}$~Hz). Figure~2 shows the
range of HBO frequencies found for Scorpius~X-1. During all
observations, the deviation from the mean $\nu_{HBO}\approx43.5$~Hz
did not exceed ${\sim5}$~Hz (i.e., it changed by no more than $10\%$
during all observations).

\begin{figure}[htb]
\epsfxsize=9.5cm
\epsffile{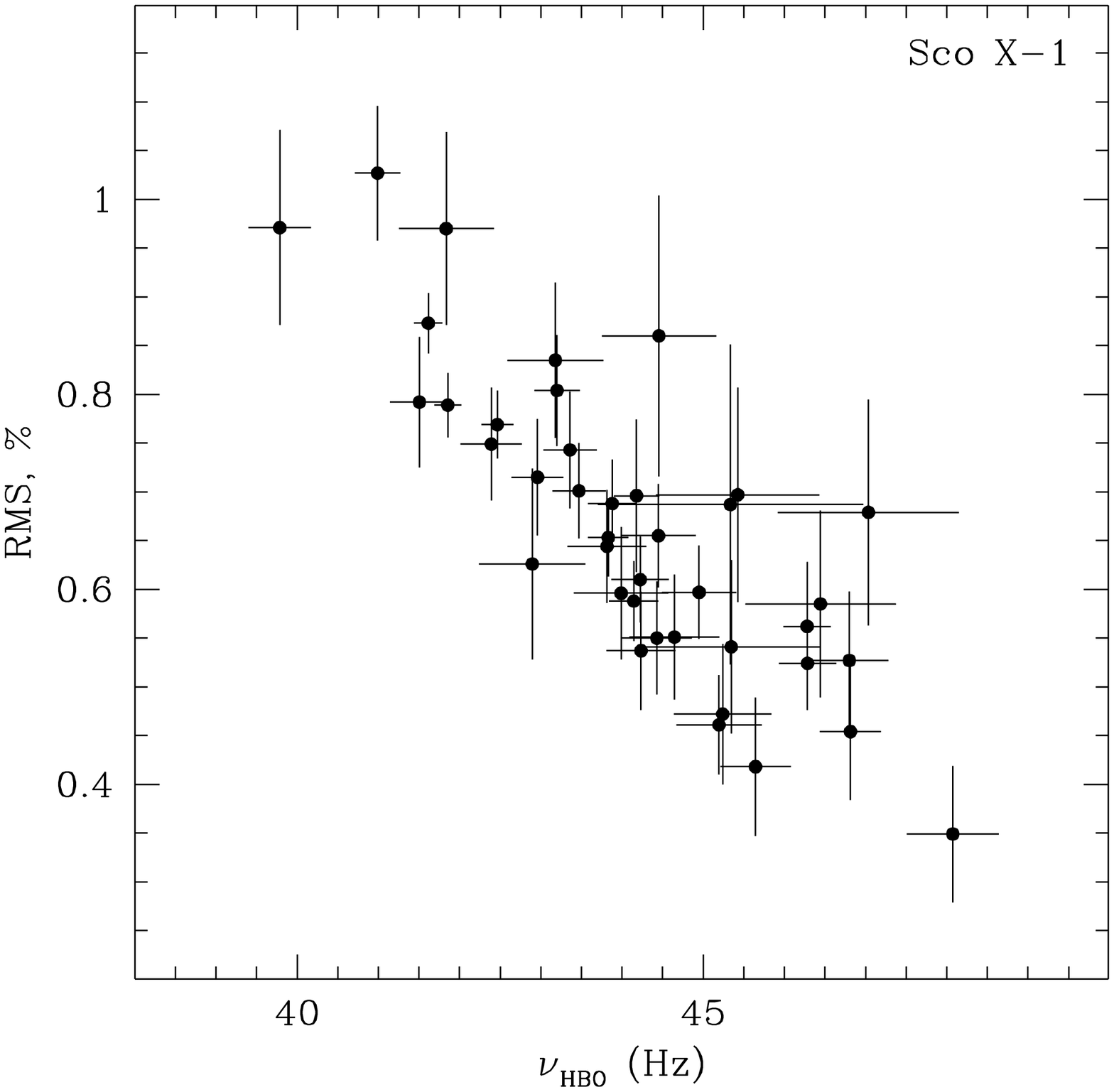}
\rput[tl]{0}(9.5,7.3){
\begin{minipage}{7.2cm}
\normalsize\parindent=0.0mm
{\small {\bf Fig.~2}. The rms of the QPO peaks versus HBO
frequency. Individual observations are presented. As
rms decreases, the detection confidence level of the
QPO peaks falls, reaching $\sim4\sigma$ for $rms$ in
the range $\sim0.4-0.5\%$.}
\end{minipage}}
\end{figure}

\subsection{Horizontal-Branch Oscillations and Their Harmonics}

In Scorpius~X-1, quasi-periodic HBO peaks were found at low
frequencies ($< 100$~Hz). Together with the oscillations at the main
frequency $\nu_{{HBO}}$, we also found QPO peaks at frequencies
close to the multiple ones. Note that the detection confidence level
of HBO peaks decreases with frequency (see Fig.~2). That is why no
multiple peaks were found (at a confidence level ${> 4\sigma}$) for
which ${\nu_{{HBO}}>48}$~Hz (the corresponding oscillation
frequency of the second harmonic is ${\nu_{2{HBO}}>96}$~Hz).

\begin{figure}[htb]
\epsfxsize=9.5cm
\epsffile{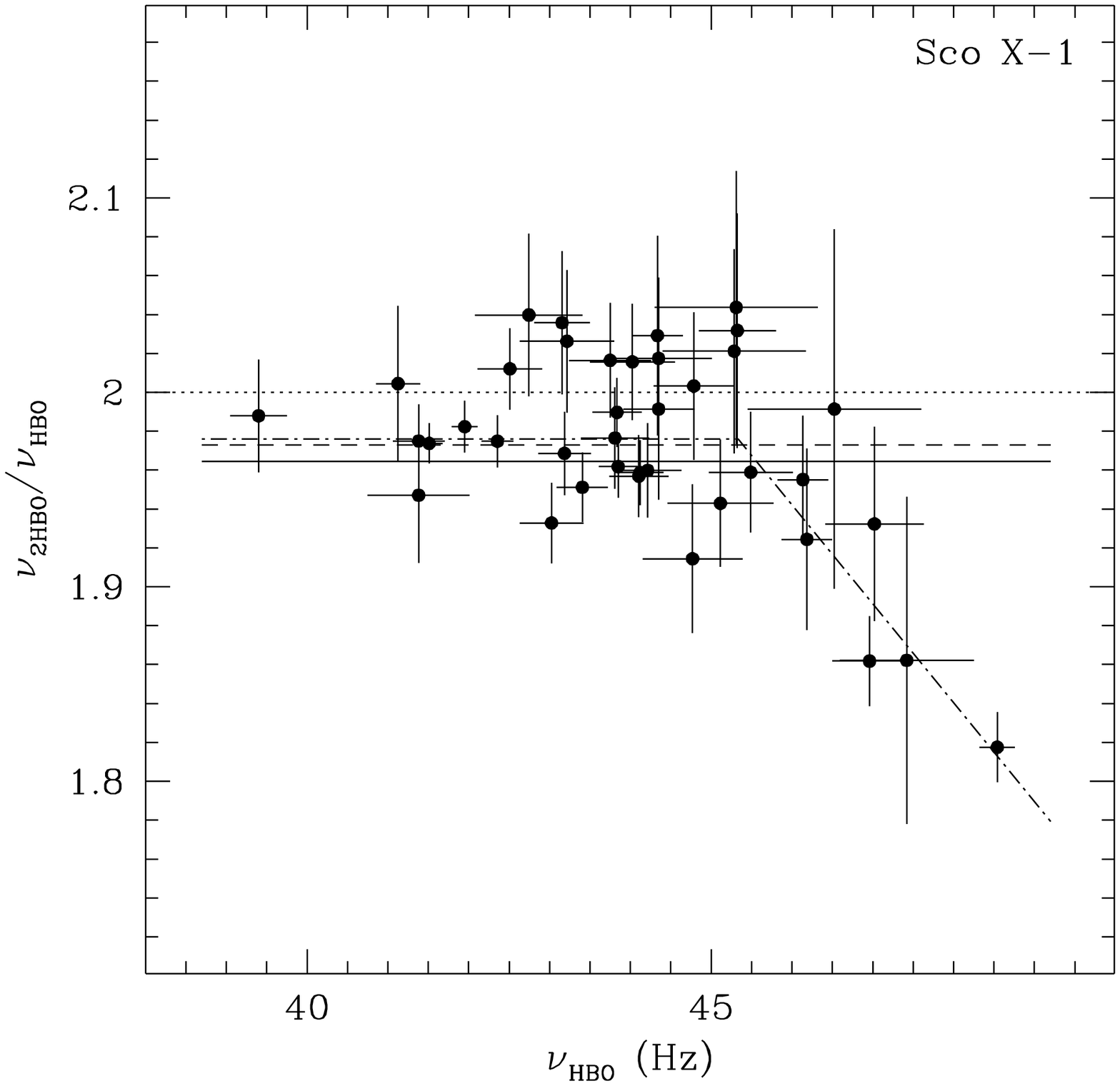}
\rput[tl]{0}(9.5,7.3){
\begin{minipage}{7.2cm}
\normalsize\parindent=0.0mm
{\small {\bf Fig.~3}. The ratio of multiple HBO harmonics versus
frequency $\nu_{{HBO}}$. The solid,
dash--dotted, dashed, and dotted lines represent,
respectively, a constant fit to the data with
statistical errors (minimization according to the
$\chi^{2}$ test), a constant fit with a break, the
mean of individual observations (without errors), and
the assumed harmonic ratio~2.0.}
\end{minipage}}
\end{figure}

The frequency ratio $r_{2/1}=\nu_{{HBO}}/\nu_{2{HBO}}$ of the multiple
HBO harmonics is shown in Fig.~3. We clearly see that $r_{2/1}$
differs from 2.0: only a third of the 39~data points are above~2.0
(the dotted line in Fig.~3). Fitting by a constant yields
$r_{2/1}=1.965\pm0.004$, $\chi^2_{{red}}=3.45$ (in what follows,
$\chi^2_{{red}}=\chi^2/{d.o.f}$ is the reduced $\chi^2$ value per
degree of freedom), while fitting by a constant with a break yields
$r_{2/1}=1.976\pm0.004$, $\chi^{2}_{{red}}=1.05$ (and
$r_{2/1}=-0.05\times(\nu_{{HBO}}-45.31)$ in the range of a linear
dependence). When the scatter of data points exceeds the stochastic
scatter and when the $\chi^2_{{red}}$ value is much larger than unity
(as in the case of a constant fit), it should be assumed that the
uncertainty could be produced by stochastic processes rather than by
the statistics alone. We can then ignore the measurement error and
determine the mean $\bar{r}_{2/1}=\frac{1}{N}\sum r_{2/1}$ and the rms
deviation from it
$\sigma=\sqrt{\frac{\sum(r_{2/1}-\bar{r}_{2/1})^2}{(N-1)}}$.  In this
case, we obtained the following ratio of the second and first
harmonics: $\bar{r}_{2/1}=1.973$, $\sigma=0.050$. Under such an
assumption, the harmonic ratio is compatible with 2.0, within the
error limits.

Let us consider in more detail the case where the scatter of $r_{2/1}$
values about the mean is determined by the statistics alone and is
incompatible with 2.0, within the error limits.  Resonances in
nonlinear oscillatory systems can emerge at frequencies at which an
external periodic force acts on the oscillator. This requires that its
frequency $\gamma$ satisfy the condition
\begin{equation}\label{eq:harm:Kuznetsov_n}
\gamma = \omega_{0}\frac{p}{q},
\end{equation}
where $p$ and $q$ are integers and $\omega_{0}$ is
the cyclic frequency of free system oscillations.

The power of the main peak must be maximal among all of the harmonics,
because the intensity of resonance phenomena rapidly decreases with
increasing $p$ and $q$ (Landau and Lifshitz~1965). In addition, if the
frequency of the periodic external force $\gamma$ differs from
$\omega_{0}$ by a small value $\varepsilon$ (in the simplest case,
${p=1}$ and ${q=1}$), then in nonlinear oscillations, this can lead to
a displacement of the rms maximum from the oscillator resonance
frequency (Landau and Lifshitz~1965). In Scorpius~X-1, the largest rms
among all the HBO harmonics corresponds to the frequency $\nu_{{HBO}}$
(see Fig.~2). We assume from the outset that the latter (and this
assumption proved to be valid) was the fundamental harmonic and the
deviation of the frequency ratio of the second and first harmonics
from 2.0 is admitted in the forced oscillations of a nonlinear
oscillator under the action of a periodic force.

Note that in this case, the frequency of the observed low-frequency
oscillation peak is $\nu_{{HBO}}\neq\omega_{0}/2\pi$. This is
particularly important in comparing the self-consistency of the
theoretical models under consideration, which interpret the emergence
of QPOs (to be more precise, the following three QPOs: $\nu_{{HBO}}$,
$\nu_{1}$, and $\nu_{2}$) in binary systems with neutron stars.

If the deviation of the ratio of the multiple HBO harmonics
$\nu_{2/1}$ from 2.0 is assumed to be attributable to forced nonlinear
oscillations, then the observed $\nu_{{HBO}}$ must be close but not
exactly equal to $\omega_{0}/2\pi$, the assumed frequencies of free
oscillations of the oscillator.  Note that $\omega_{0}/2\pi$ rather
than $\nu_{{HBO}}$ corresponds to the low-frequency QPOs in the models
considered below.

For this reason, for our analysis in terms of each of the models
considered below, we use two values for the low-frequency QPOs:
$\nu_{{HBO}}$ and $\bar{\nu}_{{HBO}}$. These values are the
fundamental HBO harmonic and the mean between half the frequency of
the second harmonic $\nu_{2{HBO}}$ and the fundamental harmonic
$\nu_{{HBO}}$. For the latter case, from all the observations, we
chose those in which the confidence level of each HBO peak was higher
than $4\sigma$.

\section{THEORETICAL MODELS}

For our analysis in terms of the two models in question, we used the
average spectra and the main frequencies: $\nu_{{HBO}}$ (as well as
$\bar{\nu}_{{HBO}} = 1/2(\nu_{{HBO}} + 1/2\nu_{2{HBO}})$, the mean
between the fundamental harmonic and half the second harmonic),
$\nu_{1}, \nu_{2}$. The following approximate asymptotic relations
between the frequencies are known from the observations of low-mass
X-ray binaries (Stella et al. 1999): (1) $\Delta\nu=\nu_{2}-\nu_{1}
\propto \nu^{2}_{2}$; and (2) $\nu_{{HBO}} \propto \nu_{1}$. These
observed features are consistent with each of the theories under
consideration.

\subsection{The Transition-Layer Model (TLM)}

We consider the motion of a clump of matter, the QPO source on the
accretion-disk surface, in a Keplerian orbit around a neutron star
(NS) in terms of this model. The magnetospheric axis is assumed to be
not aligned with the normal to the disk surface but makes an angle
$\delta$. After multiple passages through a slightly tilted
magnetosphere, the clump comes under the effect of Coriolis
forces. These forces cause the main Keplerian oscillation frequency
$\nu_{{K}}$ to split up into two oscillation modes: radial
($\nu_{{h}}$) and perpendicular to the disk plane ($\nu_{{L}}$). The
two modes are the solution of the equation for the rotation of a body
in a noninertial frame of reference [see Osherovich and Titarchuk
(1999) for more details on its derivation and solution]. In this
model, the lower and upper kHz QPO peaks correspond to the Keplerian
($\nu_{{K}}\equiv\nu_{1}$) and hybrid ($\nu_{{h}}\equiv\nu_{2}$)
frequencies, respectively. The relation between the kHz QPO peaks is
given by
\begin{equation}\label{eq:h:Kuznetsov_n}
\nu_{{h}}=[\nu_{{K}}^2+(\Omega/\pi)^2]^{1/2} ,
\end{equation}
where $\Omega$ is the angular velocity of the magnetosphere. The
oscillation mode perpendicular to the disk surface is defined as
\begin{equation}\label{eq:l:Kuznetsov_n}
\nu_{{L}}=(\Omega/\pi)(\nu_{{K}}/\nu_{{h}})\sin\delta.
\end{equation}
Since all three characteristic frequencies $\nu_{{L}}$, $\nu_{{K}}$,
and $\nu_{{h}}$ are known from observations, the sought-for
angle~$\delta$ can be calculated by using~(4):
\begin{equation}\label{eq:delta:Kuznetsov_n}
\delta=\arcsin[(\nu_{{h}}^2-\nu_{{K}}^2)^{-1/2}(\nu_{{L}}
\nu_{{h}}/\nu_{{K}})].
\end{equation}
To a first approximation, the angular velocity of the magnetosphere
$\Omega$ is constant. A more accurate equation that describes the
dependence of $\Omega$ on radius can be derived in the multipole
magnetic-field approximation (see Osherovich et al. 1984).  Assuming
the contribution of the quadrupole component to the magnetic-field
strength in the equatorial plane to be negligible and taking into
account the dipole and octupole components, the final equation for the
angular velocity can be represented as (Osherivich and Titarchuk 1999)
\begin{equation}\label{eq:omega2pi:Kuznetsov_n}
\Omega/2\pi=C_0+C_1\nu_{{K}}^{4/3}+C_2\nu_{{K}}^{8/3}+C_3\nu_{{K}}^4,
\end{equation}
with $C_2=-2(C_1C_3)^{1/2}$.

To reconstruct the magnetospheric profile for Scorpius~X-1, according
to Eq.~(3), it will suffice to have data only for the high-frequency
range of the flux variability, i.e., $\nu_{1}$ and $\nu_{2}$.
Therefore, we used all the data in which kHZ QPO peaks were found by
removing the additional condition that the main HBO peak $\nu_{{HBO}}$
and its second harmonic $\nu_{2{HBO}}$ should be detected in the
power-density spectrum at low frequencies at a confidence level above
$4\sigma$.

The angular velocity is plotted against Keplerian frequency in
Fig.~4. Using~(6), we managed to obtain the magnetospheric profile
suggested by the model under consideration with the following
parameters: $C_0=354$~Hz, $C_1=-3.54\times10^{-2}$~Hz$^{-1/3}$,
$C_2=9.99\times10^{-6}$~Hz$^{-5/3}$, and
$C_3=-7.21\times10^{-10}$~Hz$^{-3}$. In fitting the data, we
assumed~$C_2$ to be a free parameter, whereas in the model,
$C_2=-$2$(C_1C_3)^{1/2}=$1.01$\times$10$^{-5}$. The similar values of
$C_{2}$ determined by two independent methods are a weighty argument
for the approximation used for the magnetospheric profile.

\begin{figure}[htb]
\epsfxsize=9.5cm
\epsffile{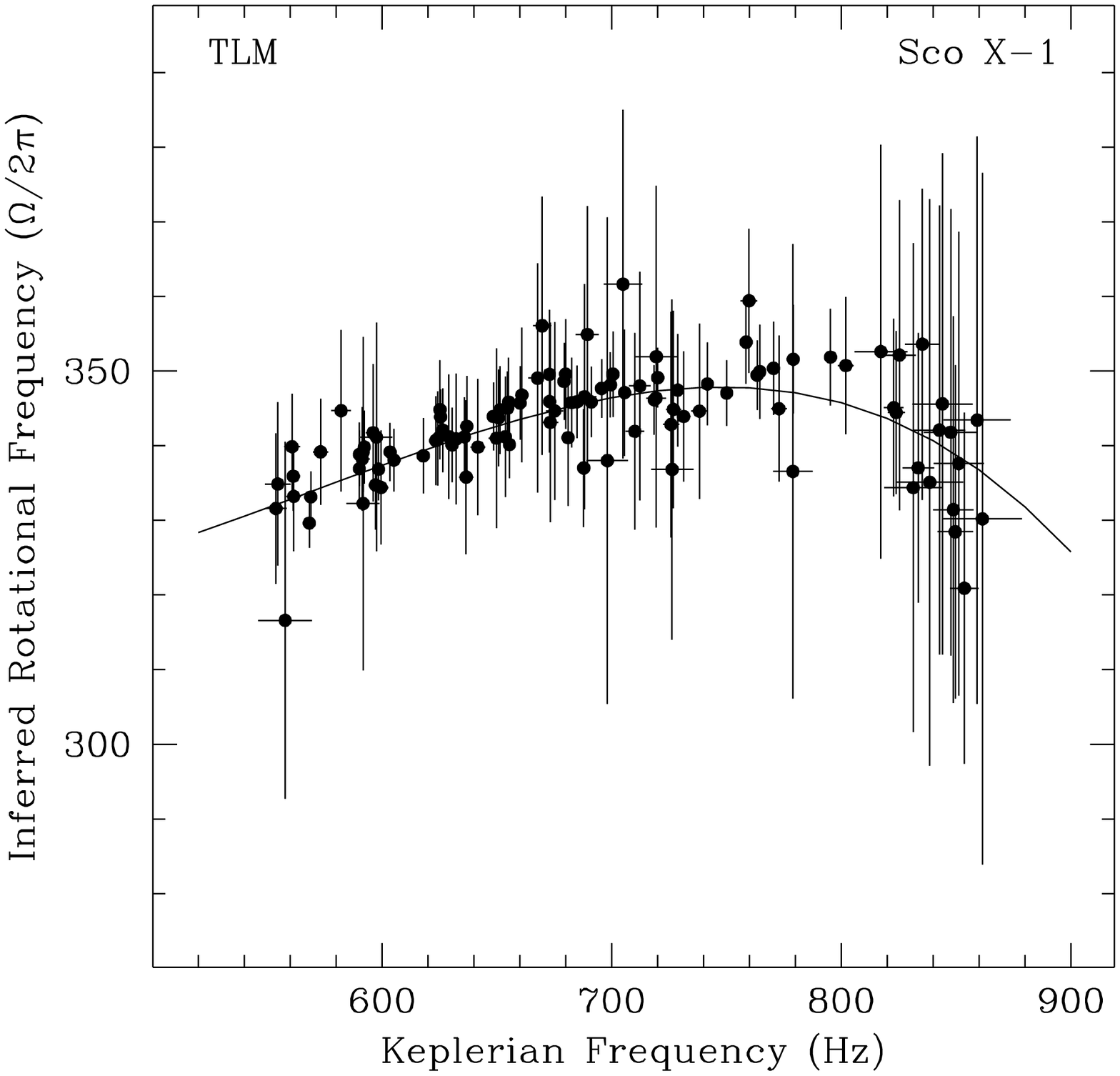}
\rput[tl]{0}(9.5,6.3){
\begin{minipage}{7.2cm}
\normalsize\parindent=0.0mm
{\small {\bf Fig.~4}. The assumed angular velocity of the
magnetosphere in the TLM, in units of $\Omega/2\pi$,
versus Keplerian frequency
($\nu_{{K}}=\nu_{1}$).}
\end{minipage}}
\end{figure}

Using~(5), we obtained~$\delta$ for each triplet of frequencies:
$\nu_{{HBO}}$, $\nu_{1}$, and $\nu_{2}$. In Fig.~5 the HBO
frequency~$\nu_{{HBO}}$ is plotted against the angle~$\delta$ between
the normal to the disk surface and the magnetospheric axis. Among all
three frequencies used in calculating $\delta$, the strongest
correlation is observed between $\delta$ and $\nu_{{HBO}}$. For this
reason, the dependence of $\delta$ on low-frequency QPOs is analyzed
in the model under consideration (below, we show that this is also
true for the relativistic precession model), because the model
invariants must be conserved, irrespective of the measured QPO
frequency range.

\begin{figure}[htb]
\epsfxsize=9.5cm
\epsffile{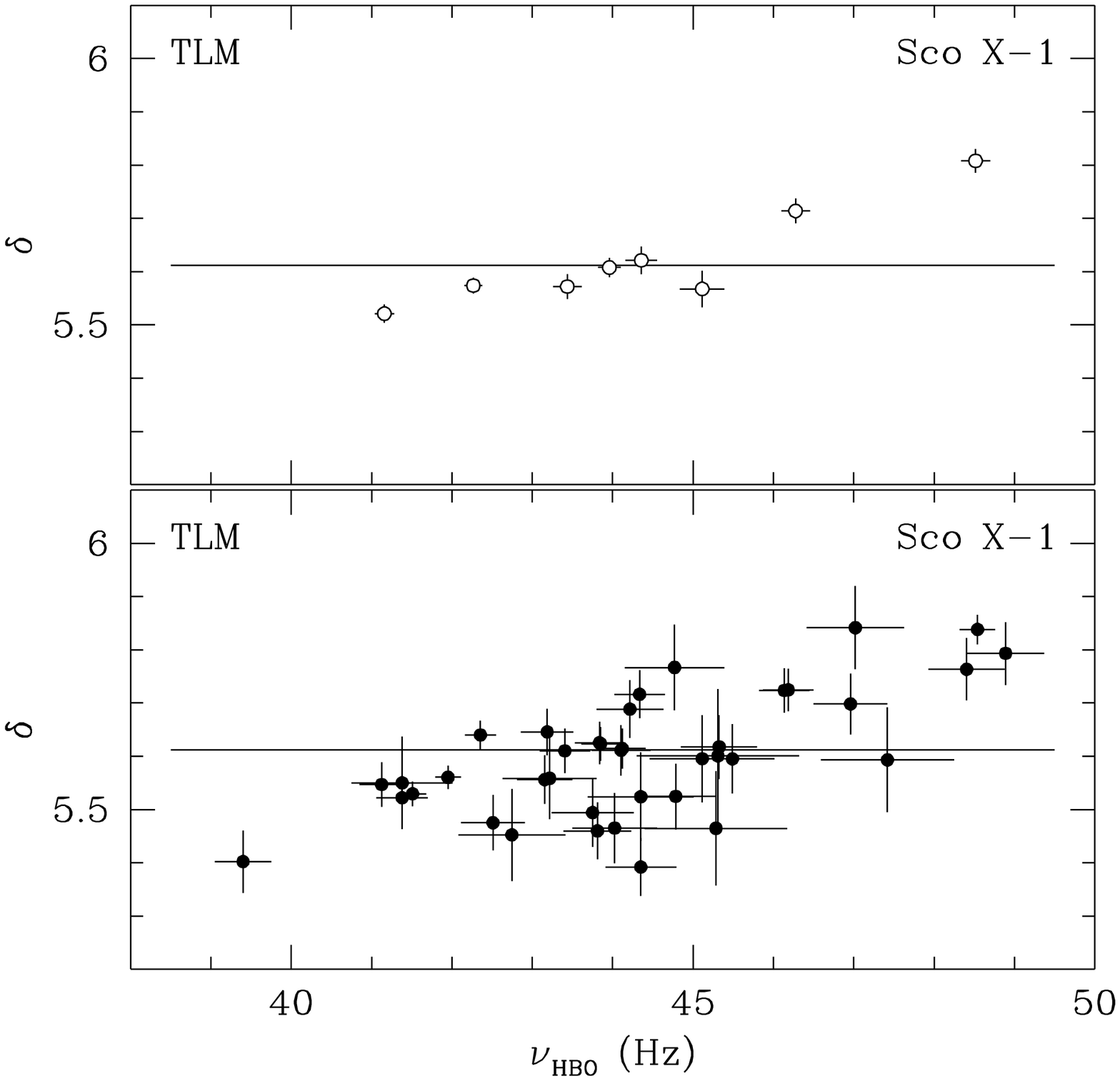}
\rput[tl]{0}(9.5,7.9){
\begin{minipage}{7.2cm}
\normalsize\parindent=0.0mm
{\small {\bf Fig.~5}. The angle between the magnetospheric axis
and the normal to the disk surface versus HBO
frequency $\nu_{{HBO}}\equiv\nu_{{L}}$.
A constant fit is indicated by the solid line. {The
lower panel} shows individual observations of
Scorpius~X-1 with a duration up to $\sim
3.5$~ks. {The upper panel} shows the same
observations as in the lower panel averaged over five
points according to the frequency
$\nu_{{HBO}}$.}
\end{minipage}}
\end{figure}

Table~1 gives data on the constant fit to~$\delta$ in two cases: when
$\nu_{{L}}=\nu_{{HBO}}$ (see Fig.~5) and
$\nu_{{L}}=\bar{\nu}_{{HBO}}=1/2(\nu_{{HBO}}+ 1/2\nu_{2{HBO}})$ (see
Fig.~6), i.e., with the inclusion of the second HBO harmonic. The
$\chi^{2}_{{red}}$ value is given in each case. The mean angle
$\bar{\delta}$ and its rms deviation $\sigma_D$ are also given in
Table.~1.

\begin{table}[htb]
\begin{center}
\begin{minipage}{7.6cm}
{\small {\bf Tabl.~1.} The angle between the normal to the disk
surface and the magnetospheric axis for various HBO
values.}
\end{minipage}\\
\begin{tabular}{l|c|c}
\hline \multicolumn{1}{c|}{Parameter} &
$\nu_{{L}}=\nu_{{HBO}}$ &
$\nu_{{L}}=\bar{\nu}_{{HBO}}^{{a}}$\\
\hline
$\delta$ & $5.612\pm0.007$ & $5.545\pm0.006$\\
$\chi^{2}_{\textrm{red}}$ & 5.53 & 2.94\\
$\bar{\delta}^{{b}}$ & $5.600\pm0.114$ &
$5.555\pm0.086$  \\
$N$, bin & 39& 39\\
\hline
\end{tabular}\\
\begin{minipage}{7.6cm}
\footnotesize{$^{{a}}\bar{\nu}_{{HBO}}=1/2(\nu_{{HBO}}
+ 1/2\nu_{2{HBO}})$.}\\
\footnotesize{$^{{b}}$The quantity
$\bar{\delta}$ is the mean of all $\delta$ with the
error $\sigma_{{D}}$ equal to the standard
deviation.}
\end{minipage}
\end{center}
\end{table}

\begin{figure}[htb]
\epsfxsize=9.5cm
\epsffile{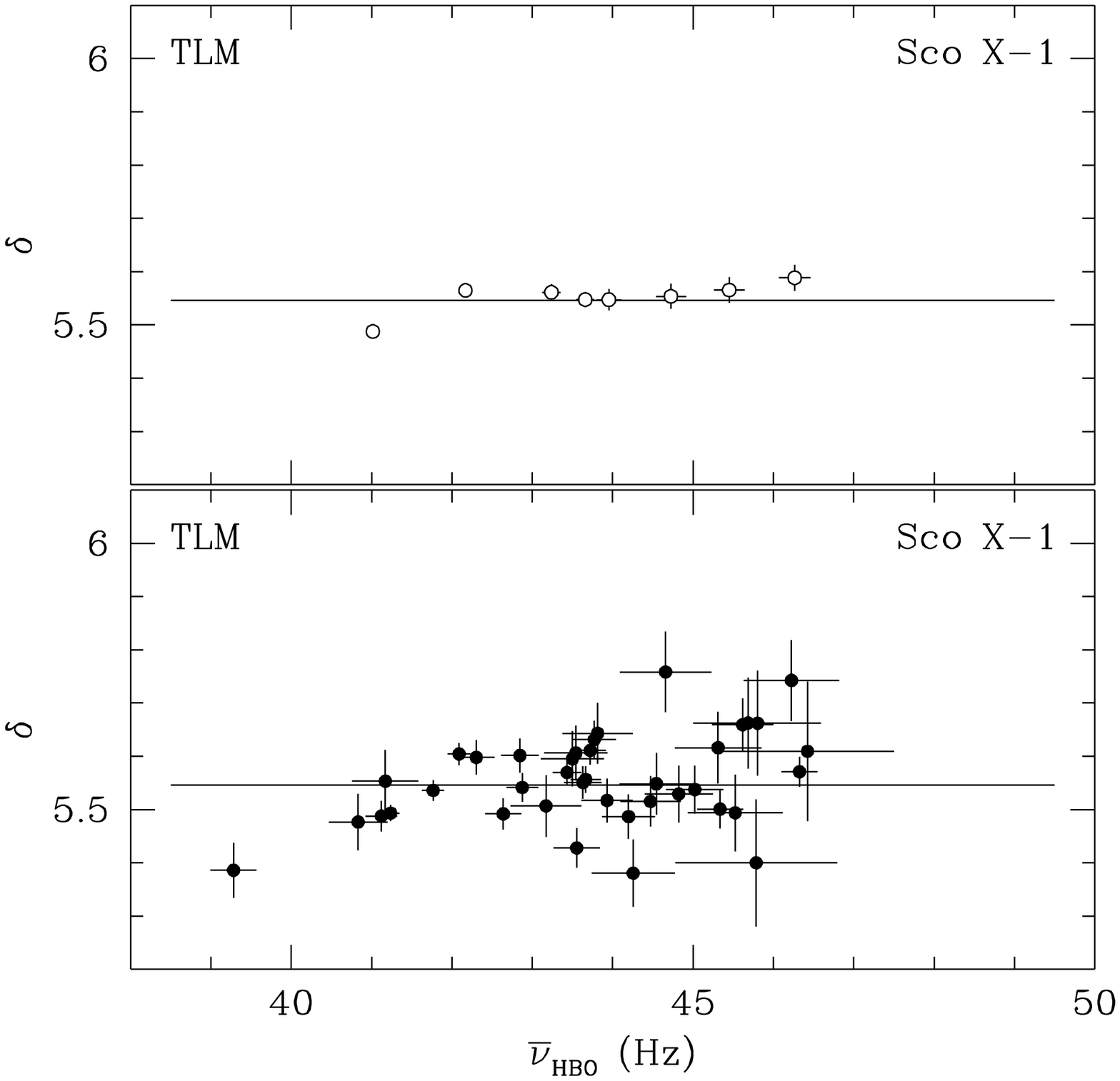}
\rput[tl]{0}(9.5,7.9){
\begin{minipage}{7.2cm}
\normalsize\parindent=0.0mm
{\small {\bf Fig.~6}. The angle between the magnetospheric axis
and the normal to the disk surface versus HBO
frequency
$\bar{\nu}_{{HBO}}\equiv\nu_{{L}}=1/2(\nu_{{HBO}}+1/2\nu_{2{HBO}})$. A
constant fit is indicated by the solid line. {The
lower panel} shows individual observations of
Scorpius~X-1 with a duration up to $\sim 3.5$~ks.
{The upper panel} shows the same
observations as in the lower panel averaged over five
points according to the frequency
$\nu_{{HBO}}$.}
\end{minipage}}
\end{figure}

To reduce the possible stochastic scatter of data points (which
probably dominates over the statistical scatter), we performed
averaging depending on the HBO frequency. The upper panels in Figs.~5
and~6 show the data obtained in this way. Whereas the data in Fig.~5
reveal a clear correlation between $\delta$ and $\nu_{{HBO}}$, the
data points in Fig.~6 virtually lie on the same line. If a larger
number of points in each bin were averaged (in our case, we combined
them by five), then ${\delta}$ (see Fig.~6, where $\delta$ is plotted
against $\bar{\nu}_{{HBO}}$) would be even more compatible with a
constant, according to the $\chi^{2}$ test.

Let us compare the rms deviation between the HBO harmonic ratio
$r_{2/1}$ determined above and $\delta$ (see Table~1). It turns out
that $\sigma_{r_{2/1}}\approx2.5\%$, while
$\sigma_{\delta}\approx2.0\%$. Since, according to~(5),
$\delta\propto\nu_{{HBO}}$ at small angles and since $\nu_{{HBO}}$
increasingly deviates from $\omega/2\pi$ to larger values with
decreasing $r_{2/1}$, the correlation for $\delta$ (clearly seen only
in Fig.~5) can result from the emergence of nonlinear oscillations
considered in the previous sections.

\subsection{The Relativistic Precession Model (RPM)}

The high-velocity motion of matter in a strong gravitational field can
generate oscillations attributable to general-relativity effects. The
RPM (see, e.g., Morsink et al.~1999) considers the motion of a point
mass around a gravitating center.  If a particle moves in an orbit
that does not lie exactly in the equatorial plane of a compact object
but is inclined at an infinitesimal angle, then the particle orbit
will precess.

For a nonrotating gravitating center (i.e., in the Schwarzschild
approximation), the expression for the particle angular velocity
matches the classical formula for Keplerian motion. However, for a NS
or a black hole with an intrinsic angular momentum, the azimuthal
frequency can be expressed (Bardeen et al. 1972; Stella et al. 1999)
in a system of units with $G=1$ and $c=1$ as
\begin{equation}\label{eq:nufi:Kuznetsov_n}
\nu_{\phi} = \sqrt{\frac{M}{r^{3}}}\hspace{-1pt}\left[
2\pi\hspace{-1pt}\left(1+a\sqrt{\frac{M}{r^{3}}}\right)\hspace{-1pt}\right]^{-1}\hspace{-1pt}.
\end{equation}
Below, we consider only the case where the compact object and the test
particle corotate: $\nu_{\phi}>0, a>0$ ($a$ is the relative angular
momentum). The epicyclic frequency $\nu_{r}$ together with the
azimuthal frequency $\nu_{\phi}$ determine the orbital periastron
rotation, $\nu_{{per}}\equiv\nu_{\phi}-\nu_{r}$, while the frequency
in a direction perpendicular to the disk plane $\nu_{\theta}$
determines the nodal precession,
$\nu_{{nod}}\equiv|\nu_{\phi}-\nu_{\theta}|$.  The corresponding
equations were derived by Okazaki et al. (1987) and Kato (1990):
\begin{equation}\label{eq:nur:Kuznetsov_n}
\nu_{r}=\nu_{\phi}\sqrt{1-6\frac{M}{r}+8a\sqrt{\frac{M}
{r^{3}}}-3\frac{a^{2}}{r^{2}}} ,
\end{equation}
\begin{equation}\label{eq:nuteta:Kuznetsov_n}
\nu_{\theta}=\nu_{\phi}\sqrt{1-4a\sqrt{\frac{M}{r^{3}}}+3\frac{a^{2}}{r^{2}}}.
\end{equation}

In contrast to the TLM, the Keplerian rotation frequency in the RPM
corresponds to the upper QPO peak, $\nu_{2}=\nu_{\phi}$, while the
observed periastron precession frequency corresponds to the lower QPO
peak, $\nu_{1}=\nu_{{per}}\equiv\nu_{\phi}-\nu_{r}$.  In the Kerr
approximation, $\nu_{\phi}\neq\nu_{\theta}$ (for the Schwarzschild
case with $a=0$, the frequencies are equal:
$\nu_{\phi}=\nu_{\theta}$), the HBOs are identified with nodal
precession: $\nu_{{nod}}\equiv|\nu_{\phi}-\nu_{\theta}|$.  Some
authors (e.g., Stella et al. 1999) believe that $\nu_{{HBO}}$ is an
even harmonic of $\nu_{nod}$. In contrast to
$\nu_{{HBO}}=\nu_{{nod}}$, this relation causes the mass of the
compact source in Eqs.~(7)--(9) required by the RPM to decrease.

In the RPM approximation under consideration, the NS mass and relative
angular momentum are invariants and do not depend on Keplerian
frequency. Thus, we can determine $M_{{NS}}$ and $a$ from
Eqs.~(7)--(9) using the three observed main frequencies:
$\nu_{{HBO}}$, $\nu_{1}$, and $\nu_{2}$.

As in the section devoted to the TLM, here, the invariant (more
specifically, the NS mass $M_{{NS}}$ suggested by the model) was also
determined in two cases: $\nu_{{nod}}=\nu_{{HBO}}$ and
$\nu_{{nod}}=\bar{\nu}_{{HBO}}=1/2(\nu_{{HBO}}+1/2\nu_{2{HBO}})$
(i.e., with the inclusion of the second HBO harmonic). Since the
derived $M_{{NS}}$ proved to be most dependent (as in the TLM) on
low-frequency HBOs, Figs.~7 and~8 show $M_{{NS}}$ as a function of
$\nu_{{HBO}}$.

\begin{figure}[htb]
\epsfxsize=9.5cm
\epsffile{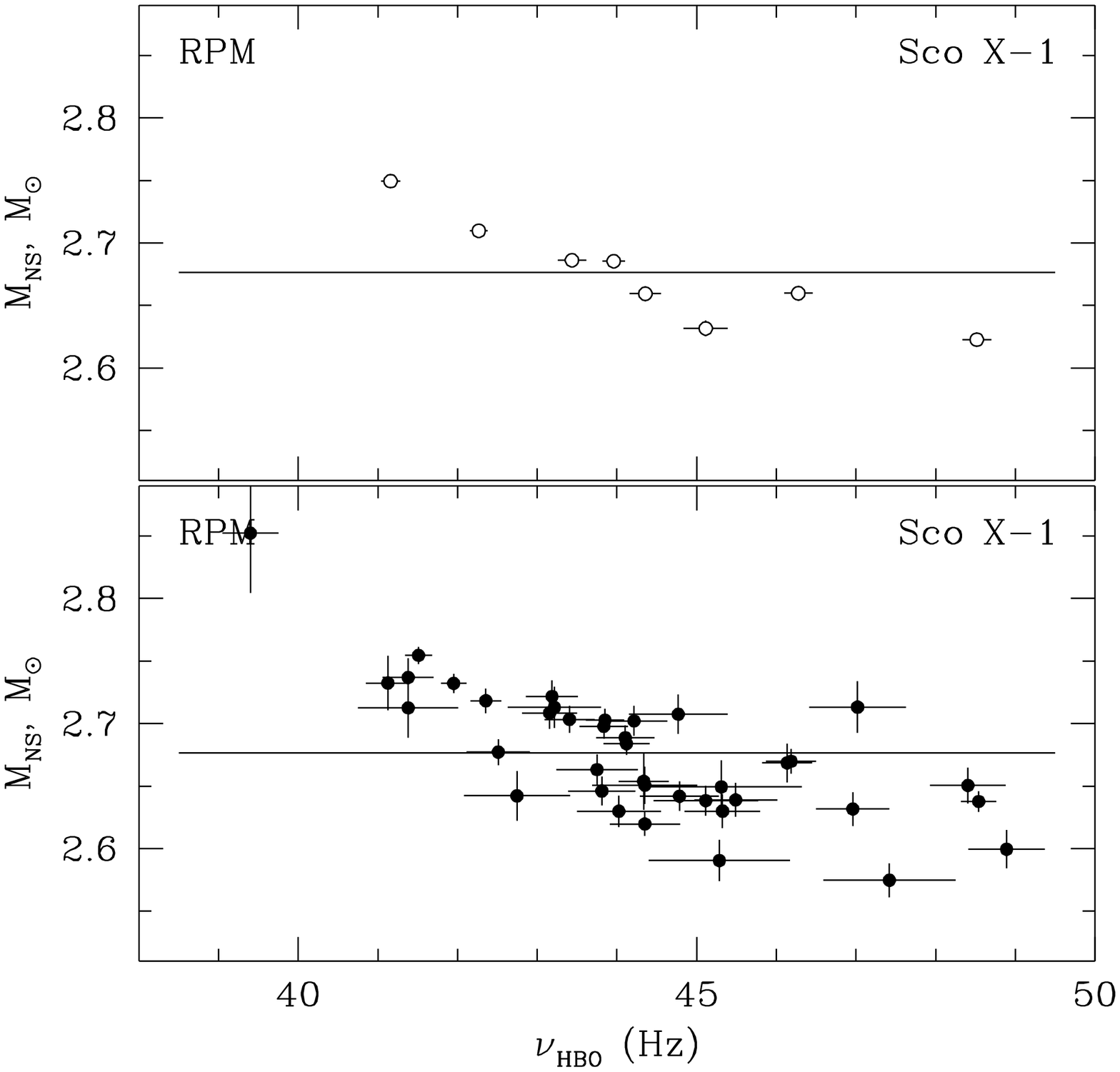}
\rput[tl]{0}(9.5,7.0){
\begin{minipage}{7.2cm}
\normalsize\parindent=0.0mm
{\small {\bf Fig.~7}. NS mass versus observed HBO frequency
$\nu_{{HBO}}$. The solid line indicates a
constant fit to the data. {The lower panel} shows
individual observations of Scorpius~X-1 with a
duration up to $\sim 3.5$~ks. {The upper panel} shows
the same observations as in the lower panel averaged
over five points according to the frequency
$\nu_{{HBO}}$.}
\end{minipage}}
\end{figure}

\begin{table}[htb]
\begin{center}
\begin{minipage}{12.8cm}
{\small {\bf Tabl.~2.}The NS mass and its relative angular momentum for various HBO values.}
\end{minipage}
\begin{tabular}{l|c|c|c|c}
\hline \multicolumn{1}{c|}{Parameter} &
$\nu_{{nod}}=\nu_{{HBO}}$$^{{a}}$ &
$\nu_{{nod}}=\bar{\nu}_{{HBO}}$ &
$2\nu_{{nod}}=\nu_{{HBO}}$ &
$2\nu_{{nod}}=\bar{\nu}_{{HBO}}$$^{{a}}$\\
\hline
$M_{{NS}}, M_{\odot}$ & $2.680\pm0.002$ & $2.675\pm0.002$&$2.339\pm0.002$ & $2.338\pm0.002$\\
$\chi^{2}_{{red}}$ & 14.3  & 16.3& 16.4 & 17.5\\
$\bar{M}_{{NS}}^{{b}}, M_{\odot}$ & $2.677\pm0.050$ & $2.674\pm0.053$ & $2.341\pm0.045$ & $2.341\pm0.045$ \\
$a^{{c}}$, ¬ & $1367\pm2$\phantom{.000} & $1358\pm2$\phantom{.000}& $663\pm1$\phantom{.000}& $658\pm1$\phantom{.000}\\
$\chi^{2}_{{red}}$ &8.2 & 13.7 &8.2 & 13.6\\
$N$, bin & 39& 39& 39& 39\\
\hline
\end{tabular}\\
\begin{minipage}{12.8cm}
\footnotesize{$^{{a}}\bar{\nu}_{{HBO}}=1/2 (\nu_{{HBO}} + 1/2\nu_{2{HBO}})$.\\
$^{{b}}\bar{M}_{{NS}}$ corresponds to the mean of all ${M}_{{NS}}$
with the error equal to the standard deviation $\sigma_D$.\\
$^{{c}}a=J/(c\times M_{{NS}})$.}
\end{minipage}
\end{center}
\end{table}

Table~2 gives a constant fit to the data (shown in Figs.~7 and.~8). We
clearly see a large $\chi^{2}_{{red}}$ value for each $\nu_{{nod}}$:
compared to Table.~1, which gives a fit to $\delta$ (in the TLM), the
corresponding $\chi^{2}_{{red}}$ values are larger by a factor of
$\sim3-6$.  Nevertheless, the rms deviation $\sigma_{M_{{NS}}}$ is
also $\approx2.0\%$, as is $\sigma_{\delta}$ in the TLM. In contrast
to Fig.~4, in which $\delta$ is plotted against $\bar{\nu}_{{HBO}}$
and is essentially compatible with a constant, we clearly see a linear
correlation of $M_{{NS}}$ with $\bar{\nu}_{{HBO}}$ in Fig.~8: as
$\bar{\nu}_{{HBO}}$ increases from $\sim39.5$~Hz to $\sim46.5$~Hz, the
required $M_{{NS}}$ decreases from $\sim2.8M_{\odot}$ to
$\sim2.6M_{\odot}$.

\begin{figure}[htb]
\epsfxsize=9.5cm
\epsffile{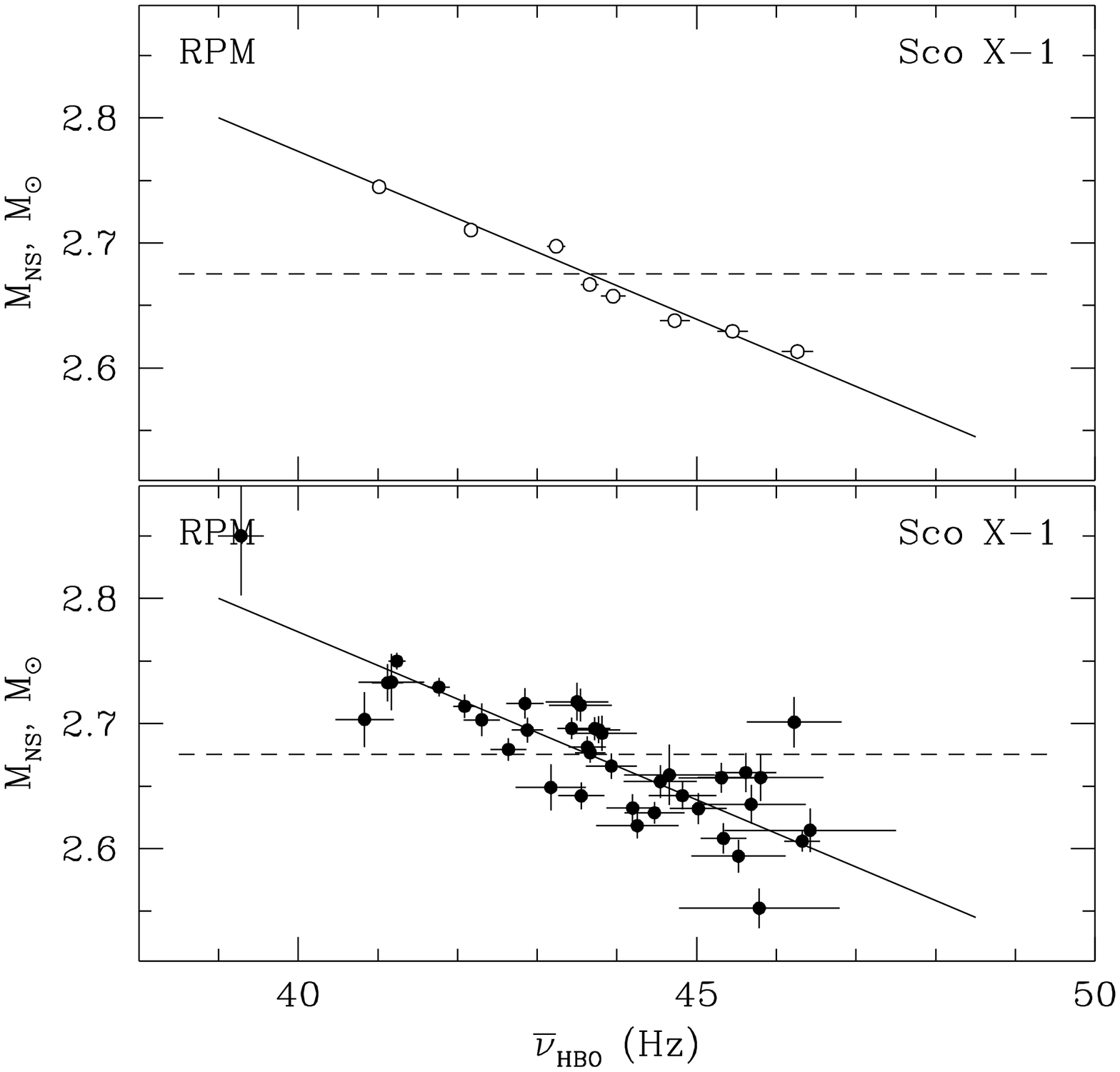}
\rput[tl]{0}(9.5,7.6){
\begin{minipage}{7.2cm}
\normalsize\parindent=0.0mm
{\small {\bf Fig.~8}. NS mass versus observed HBO frequency
$\nu_{{nod}}\equiv\bar{\nu}_{m{HBO}}=1/2 (\nu_{{HBO}} +
1/2\nu_{2{HBO}})$. The solid line indicates a constant fit to the
data. {The lower panel} shows individual observations of
Scorpius~X-1 with a duration up to $\sim 3.5$~ks. {The upper
panel} shows the same observations as in the lower panel averaged over
five points according to the frequency $\nu_{{HBO}}$.}
\end{minipage}}
\end{figure}

The relativistic precession model makes it possible to estimate the
compact-source mass and to determine the Keplerian orbital radius. For
comparison, Fig.~9 shows the mass--radius relations for various
equations of state for neutron stars taken from Miller et al.~(1998).

\begin{figure}[htb]
\epsfxsize=9.5cm
\epsffile{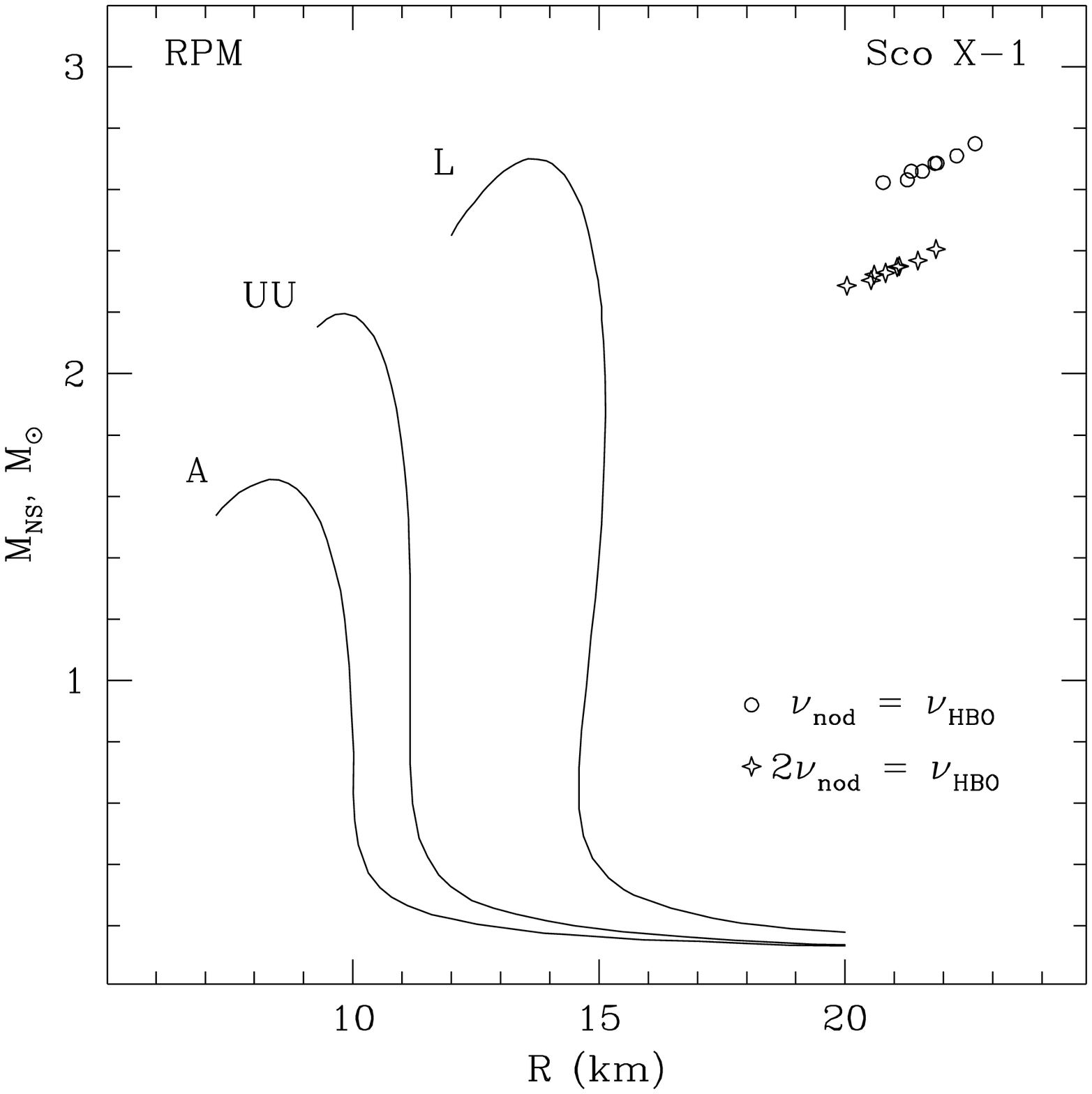}
\rput[tl]{0}(9.5,8.5){
\begin{minipage}{7.2cm}
\normalsize\parindent=0.0mm
{\small {\bf Fig.~9}. Relationship of the NS mass to the nodal
precession frequency and Keplerian orbital radius.
The circles correspond to the mass--radius relation
in the case where the nodal precession frequency is
identically equal to the HBO frequency
$\nu_{{HBO}}$; the diamonds correspond to the
case where it is assumed to be its even harmonic (for
more details, see the text). The solid lines [taken
from Miller et al.~(1998)] represent the
mass-radius relations for several equations of state
for neutron stars.}
\end{minipage}}
\end{figure}

\subsection{Comparison between the Assumed Invariants of the
Models under Consideration}

\begin{table}[htb]
\begin{center}
\begin{minipage}{12.0cm}
{\small {\bf Tabl.~3. }Correlation between model invariants,
$\delta$ in the TLM and $M_{{NS}}$ in the RPM,
and HBO frequency.}
\end{minipage}
\begin{tabular}{l|c|c|c|c}
\hline
 & \multicolumn{4}{c}{Model}\\ \cline{2-5}
\multicolumn{1}{c|}{Parameter} &
\multicolumn{2}{c|}{TLM} & \multicolumn{2}{c}{RPM}\\
\cline{2-5}
 & $\nu_{{L}}=\nu_{{HBO}}$ &
 $\nu_{{L}}=\bar{\nu}_{{HBO}}$$^{{a}}$&
  $\nu_{{nod}}=\nu_{{HBO}}$ &
   $\nu_{{nod}}=\bar{\nu}_{{HBO}}$$^{{a}}$\\
\hline
$r_{s}$ & 0.567& 0.359 & --0.715 & --0.789\\
Prob &  $2\times10^{-4}$ & $3\times10^{-2}$ & $3\times10^{-7}$ & $2\times10^{-9}$ \\
$N$, bin & 39& 39& 39& 39\\
\hline
\end{tabular}\\
\begin{minipage}{12.0cm}
\footnotesize{$^{{a}}\bar{\nu}_{{HBO}}=1/2
(\nu_{{HBO}} + 1/2\nu_{2{HBO}})$\\
Note. We used the individual observations shown in the
lower panels of Figs.~5-8. The Spearman
coefficient~$r_{s}$ close to $1$ or $-1$ corresponds
to a total positive or negative correlation between
the two quantities. The probability that the observed
correlation is the result of purely statistical
fluctuations is given for each case.\\ \\ 
{\bf Tabl.~4. } Correlation between model invariants,
$\delta$ in the TLM and $M_{{NS}}$ in the RPM,
and HBO frequency.} 
\end{minipage}
\begin{tabular}{l|c|c|c|c}
\hline & \multicolumn{4}{c}{Model}\\ \cline{2-5}
\multicolumn{1}{c|}{Parameter} &
\multicolumn{2}{c|}{TLM} & \multicolumn{2}{c}{RPM}\\
\cline{2-5} & $\nu_{{L}}=\nu_{{HBO}}$ &
$\nu_{{L}}=\bar{\nu}_{{HBO}}$
$^{{a}}$&
$\nu_{{nod}}=\nu_{{HBO}}$ &
$\nu_{{nod}}=\bar{\nu}_{{HBO}}^{{a}}$\\
\hline
$r_{s}$ & 0.738 & 0.595 & --0.929 & --1.000\\
Prob &  $3.7\times10^{-2}$& $1.2\times10^{-1}$& $8.6\times10^{-4}$& 0.0\\
$N$, bin & 8 & 8 & 8 & 8\\
\hline
\end{tabular}\\
\begin{minipage}{12.0cm}
\footnotesize{$^{{a}}\bar{\nu}_{{HBO}}=1/2
(\nu_{{HBO}} + 1/2\nu_{2{HBO}})$\\ 
Note. We used the data (averaged over four or five observations) shown
in the upper panels of Figs.~5-8. The Spearman coefficient~$r_{s}$
close to $1$ or $-1$ corresponds to a total positive or negative
correlation between the two quantities. The probability that the
observed correlation is the result of purely statistical fluctuations
is given for each case.}
\end{minipage}
\end{center}
\end{table}

To check whether the two models are self-consistent and to determine
whether the derived invariants $\delta$ and $M_{{NS}}$ correlate with
the HBO frequency (see Fig.~5--9), we used Spearman's nonparametric
correlation test. The coefficient~$r_{s}$ equal to unity in magnitude
points to a total correlation between the two quantities. In each case
under consideration, we give the probability~(prob) that the
correlation found is the result of statistical fluctuations. This test
is also convenient in the case where it is necessary to establish
which of the several correlations under consideration is
strongest. Our estimates are given in Tables~3 and~4. Note that the
data in Table~4 correspond to the case where the statistical scatter
was slightly reduced through averaging. Thus, the RPM model (see
Table~4) is inconsistent (${r_{s}=-1}$; the total negative correlation
between $M_{{NS}}$ and $\bar{\nu}_{{HBO}}$) and the assumed NS mass is
not an invariant in this model.

Allowance for the second harmonic in determining the low-frequency QPO
peak (more specifically,
$\bar{\nu}_{{HBO}}=1/2(\nu_{{HBO}}+1/2\nu_{2{HBO}})$) results in a
reduction of the correlation coefficient $r_{s}$ for $\delta$ in the
TLM, while the reverse is true for $M_{{NS}}$. In particular, we see
from Fig.~8 that the data points are well fitted by a straight line:
$\chi^{2}_{{red}}=4.31$ (see Table~2, where $\chi^{2}_{{red}}=16.3$
for the same case but for a constant fit to the data).  Figures~6
and~8 (upper panels) show that averaging can reduce the stochastic
scatter when determining the model invariants.  However, for the TLM
(see Fig.~6), $\delta$ is virtually constant, $\bar{\delta}=5.55$,
with the rms deviation $\sigma_{\bar{\delta}}=0.03$ (which is
$\approx0.5\%$), while for the RPM (see Fig.~8), $M_{{NS}}$ strongly
correlates with $\bar{\nu}_{{HBO}}$ and is incompatible with a
constant (the rms deviation is $\sigma_{\bar{M}_{{NS}}}\approx1.7\%$,
which roughly corresponds to $\sigma_{{M}_{{NS}}}$; see Table~2).

\section{DISCUSSION AND RESULTS}

Both theoretical models suggest the existence of invariants. These
invariants are the tilt of the magnetosphere to the disk normal
$\delta$ for the TLM and the NS mass $M_{{NS}}$ and its relative
angular momentum $a$ for the RPM in the Kerr approximation. Since the
RPM did not consider the quadrupole moment of the compact source and
disregarded its oblateness (which can give a contribution of $\sim
10-15\%$ to the nodal precession frequency), we managed to avoid the
inclusion of various theoretical equations of state for neutron stars
in our analysis. However, it was of interest to compare our results
with the constraints imposed on the equations of state. Figure~9 shows
some of these constraints taken from Miller et al.~(1998). The curves
correspond to the mass--radius relation for nonrotating neutron stars.
Although the angular velocity for Scorpius~X-1 is nonzero, the
correction for spin (here, $\nu_{s}\approx200-400$~Hz) can be
disregarded.

The assumed NS mass $M_{{NS}}$ and the Keplerian orbital radius $R$ of
the particle rotating around it (see Fig.~9) derived from
Eqs.~(7)--(9) must be consistent with the constraints imposed by the
equations of state. More specifically, ${M_{{NS}}<M_{{EOS}}}$,
${R>R_{{EOS}}}$ (here, $M_{{EOS}}$ and $R_{{EOS}}$ are the NS mass and
radius for various equations of state). Only the equation of state
``L'' suggest that a massive NS (in our case, $\sim2.7M_{\odot}$)
could exist and it agrees with the derived $M_{{NS}}$ in the two cases
($\nu_{{nod}}=\nu_{{HBO}}$ and 2$\nu_{{nod}}=\nu_{{HBO}}$). At the
same time, the other two equations of state (``A'' and ``UU'') are
incompatible with the RPM results in the Kerr (i.e., with NS rotation)
approximation.  Moreover, we obtained $M_{{NS}}$ for Cygnus~X-2 that
was compatible, within the error limits ($M_{{NS}}=2.69\pm0.10$ for
$\nu_{{nod}}=\nu_{{HBO}}$ and $M_{{NS}}=2.29\pm0.09$ for
$2\nu_{{nod}}=\nu_{{HBO}}$; Kuznetsov~2002), with $M_{{NS}}$ for
Scorpius~X-1 (see Table~2). Such a coincidence is not an argument for
the relativistic precession model either.

In addition, we showed that a nonparametric test for correlation
between the assumed NS mass in the RPM and the HBO frequency yields a
positive result. Thus, the RPM is not a self-consistent model in which
the invariants are conserved and the derived mass of the compact
object is large enough for the existing equations of state for neutron
stars. At the same time, for the transition-layer model, the derived
value is compatible with a constant under certain conditions (e.g.,
using the theory of nonlinear oscillations; for more details, see
Kuznetsov and Titarchuk~2002). If the uncertainty in the calculated
angle $\delta$ does not exceed $\sim2\%$ (which is lower than the rms
deviation for $\delta$ in Scorpius~X-1; see Table~1), then its value
may be assumed to be constant.

The observed inconsistency of the RPM may stem from the fact that the
orbital inclination of the test particle in the model simplification
is assumed to be infinitesimal. An exact solution for an arbitrary
inclination is given in Sibgatullin~(2001). In such an analysis, the
frequency $\nu_{{HBO}}$ for fixed mass and angular momentum can change
by a factor of $\sim3$ as the inclination changes from 0 to $\pi/2$
(for the marginally stable orbit at $M_{{NS}}=2M_{\odot}$ and
$\nu_{\phi}=1200$~Hz). This dispenses with the need to take the nodal
precession frequency $\nu_{{nod}}$ to be equal to 1/2 of $\nu_{{HBO}}$
to obtain more acceptable NS-star masses.

In addition, it is worth noting that the angle $\delta$ in the RPM may
not be strictly an invariant.  Higher energy release in the disk can
affect the curvature of its inner region. The tilt angle $\delta$
between the magnetospheric axis and the normal to the disk surface can
then differ for different source fluxes. As a result, the angles
$\delta$ can differ for the same $\nu_{{HBO}}$ (see Figs.~5 and 6).

\section*{ACKNOWLEDGMENTS}

This study was supported in part by the Program of the Russian Academy
of Sciences ``Astronomy: Nonstationary Astronomical Phenomena'', the
Russian Foundation for Basic Research (project no.~02-02-17347), and
grant no.~00-15-99297 of the President of Russia. I used the RXTE data
retrieved from the Goddard Space Flight Center Electronic Archive. I
wish to thank L.~Titarchuk, B.~Stone, N.~White,
J.~Swank, Ph.~Newman, and J.~Repaci for the opportunity to work with
the RXTE archival data on CR-ROMs. I am grateful to the referee for
helpful remarks and comments.

Translated by V. Astakhov

\end{document}